# Continuously Red-Shift and Blue-Shift Wavelength-Tuneable, Narrowband, High Harmonics in the EUV – X-ray Regime for Resonance Imaging and Spectroscopies

Dimitar Popmintchev,[1,3] Aref Imani,[1] Paolo Carpegiani,[1] Joris Roman,[1] Siyang Wang,[2] Jieyu Yan,[2] Sirius Song,[2] Ryan Clairmont,[2] Zhihan Wu,[2] Elizaveta Gangrskaia,[1] Edgar Kaksis,[1] Tobias Flöry,[1] Audrius Pugžlys,[1] Andrius Baltuška,[1] Tenio Popmintchev[1,2,4]

[1]*Photonics Institute, TU Wien, Vienna A-1040, Austria*
[2]*University of California San Diego, Physics Department, Center for Advanced Nanoscience, La Jolla, CA 92093, USA*
[3]dimitar.popmintchev@tuwien.ac.at
[4]tenio.popmintchev@physics.ucsd.edu

**We demonstrate a novel technique for producing high-order harmonics with designer spectral combs in the extreme ultraviolet–soft X-ray range for resonance applications using spectrally controlled visible lasers. Our approach enables continuous tunability of the harmonic peaks while maintaining superb laser-like features such as coherence, narrow bandwidth, and brightness. The harmonics are conveniently shifted towards lower or higher energies by varying the infrared pulse parameters, second harmonic generation phase-matching conditions, and gas density inside a spectral-broadening waveguide. In the time domain, the X-rays are estimated to emerge as a train of sub-300 attosecond pulses, making this source ideal for studying dynamic processes in ferromagnetic nanostructures and other materials through resonant multidimensional coherent diffractive imaging or other X-ray absorption spectroscopy techniques. Moreover, the visible driving laser beams exhibit an ultrashort sub-10 fs pulse dues to nonlinear self-compression with a more than 30-fold enhancement in peak intensity that also extends the tunability of the linewidth of the harmonic combs.**

**Introduction.** State-of-the-art tabletop X-ray sources based on the process of high-order harmonic generation can deliver ultrafast coherent light with high brightness, unrivaled spatial coherence, and attosecond scale temporal structure [1-7]. They have an exceptional ability to penetrate opaque objects, visualize subwavelength nanostructures, control spin, and magnetic currents, access the chemistry and structure of materials at a picometer length and femtosecond-to-attosecond temporal scales, etc. [7-12]. Nonetheless, the wavelength tunability of these light sources remains, thus far, very limited. The importance of this feature for tabletop sources has been anticipated to access the dimension of the material's identity through their absorption-edge fingerprints. Most of the current understanding of matter's atomic or molecular structure and its dynamics is based on diffraction measurements, or more specifically, on the interaction of the light's electric field with the charge density of matter. On the other hand, the magnetic scattering cross-section appears as a small relativistic correction. Even at resonance, the amplitude of the resulting magnetic diffraction can be orders of magnitude smaller than that of the charge scattering amplitude. Accordingly, resonant magnetic diffraction imaging has been one of the most challenging experiments considering that narrow bandwidth, near-edge, ultrabright X-ray light is a prerequisite in this case. Here we demonstrate a fully spatially and temporally coherent, continuously wavelength-tunable, extreme ultra-violet (EUV) – soft X-ray source covering ferromagnetic *N* and *M* absorption edges. More generally, the material absorption edges encode the instantaneous charge and spin state localization to a particular element or functional group, giving supplementary information about the local structure, chemical nature, and electronic structure, as well as orbital and spin ordering phenomena. This data is accessible through the X-ray absorption fine-structure spectroscopies (XAFS), which are sensitive to the electronic, spin, and geometric structure of molecules and materials [12]. These techniques are complementary to optical spectroscopies, which can reveal the non-equilibrium dynamics of electronic or magnetic phenomena at ultrafast time scales. A tunable source is thus ideal for such experimental implementations, as it features well-spaced harmonics - $4.8 - 5.2\ eV$ - with excellent peak-to-valley contrast, allowing harmonic isolation with EUV mirrors or monochromators and straightforward experimental implementation. Our approach combines the continuous tunability of the harmonics towards lower or higher wavelengths in the EUV to the soft X-ray regimes covering the entire photon energy spacing between harmonics for the first time, with high brightness and narrow bandwidth of near 1 eV. The temporal structure is predicted to consist of attosecond pulse trains with a characteristically low attosecond chirp. We refer the reader to another work for details of the VIS-driving laser source [13].

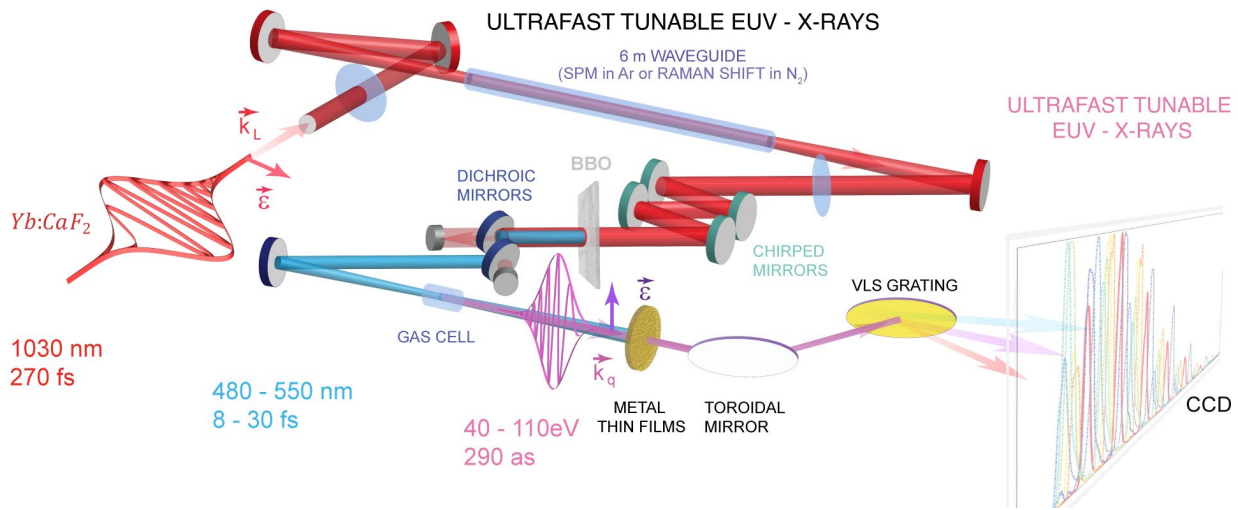

**Figure 1. Infrared pulse compression and generation of tunable VIS driving laser light at 480 – 550 $nm$.** Experimental setup: sub-picosecond pulses at 1030 $nm$ are spectrally broadened and compressed to femtosecond scales of $15 - 30\ fs$, using a gas-filled hollow-core fiber and chirped mirrors. Further, a 300 $\mu m$ long $BBO$ crystal generates the visible laser beam with a temporal duration of $8 - 30\ fs$. The pressure and the nature of the atomic or molecular gas species inside the waveguide control the spectral broadening and the spectral blue or red shift, respectively.

**Wavelength-tunable, ultrafast, coherent X-ray light.** The high-order harmonic sources have established the status quo of the tabletop attosecond science owing to the Ti:Sapphire laser amplifiers delivering beams with excellent spatial profile quality and ultrafast pulses down to the near single-cycle regime. Nonetheless, the spectral region of this laser's wavelength engenders some adverse effects on extending the cutoff energies for high harmonic generation (HHG) and on the ability to phase-match far-infrared (far-IR) frequencies in optical parametric implementations. In our experiments, we used a $Yb:CaF_2$ laser amplifier at 1030 $nm$, delivering sub-picosecond laser pulses ($\sim$240 $fs$) with up to 14 $mJ$ at $0.5 - 1\ kHz$ (Fig. 1) [14]. The pulses are spectrally broadened in a 6-meter-long hollow-core fiber with a 1 $mm$ core diameter and compressed using a set of four chirped mirrors, balancing a group delay dispersion ($GDD$) of near $600\ fs^2$ [15-22]. The broadened infra-red spectra towards the blue and red spectral regions can be produced using atomic gasses (i.e., $Ar$, $Kr$, $Xe$, etc.,) and molecular gasses (i.e., $N_2$, air, etc.), respectively. Using such an approach, the $Yb:CaF_2$ laser pulses can be compressed to sub-15 $fs$ with uncompensated higher chromatic dispersion orders. Subsequently, we use a $BBO$ type I crystal with $10 - 30\%$ conversion efficiency to obtain tunable driving pulses in the visible $480 - 550\ nm$ spectral range, with ultrashort short pulse durations of $8 - 30\ fs$, when using crystals (Fig. 2). Furthermore, we vary the infrared pulse dispersion, gas species and tune the pressure inside the waveguide. This is needed because the $BBO$ crystals can only phase-match a finite bandwidth. In such a way, we can extract a second harmonic with maximum energy and desired spectrum ideal for high harmonic upconversion. The fiber is filled with $Ar$ or $N_2$ gas at pressures $p = 30 - 1000\ mbar$ providing extreme tunability of the blue or red-shifted spectral peak of the second harmonic beams.

Interestingly, intense EUV, UV, and VIS lasers are ideal for harmonic generation since they combine favorable single-atom quantum and macroscopic phase-matching physics. Generating high harmonics by short-wavelength lasers exhibits numerous benefits - most notably, extremely high conversion efficiency, ultra-narrow bandwidth harmonics with significant energy separation, extended X-ray cutoffs, inherently naturally compressed attosecond pulse structure, effective phase and group velocity matching, increased tolerance to phase-mismatch, excellent spatio-temporal coherence, etc. [7, 23, 24].

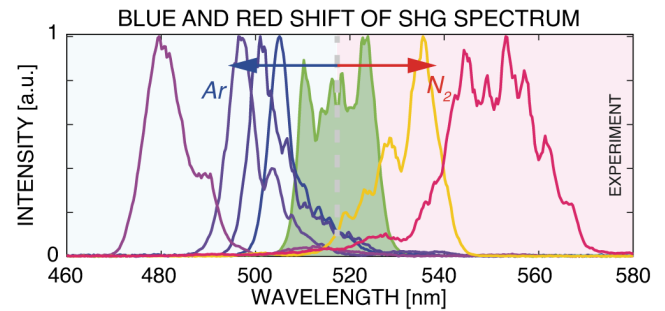

**Figure 2. Ultrafast VIS driving laser light with extreme blue and red shift tunability.** Second-harmonic spectra with tunable central wavelengths from 480 to 550 nm, generated in a BBO crystal from gas-broadened pulses in $Ar$ (blue) or $N_2$ (red) waveguide. All broadband pulses are optimized for high energy and short pulse durations for efficient HHG generation.

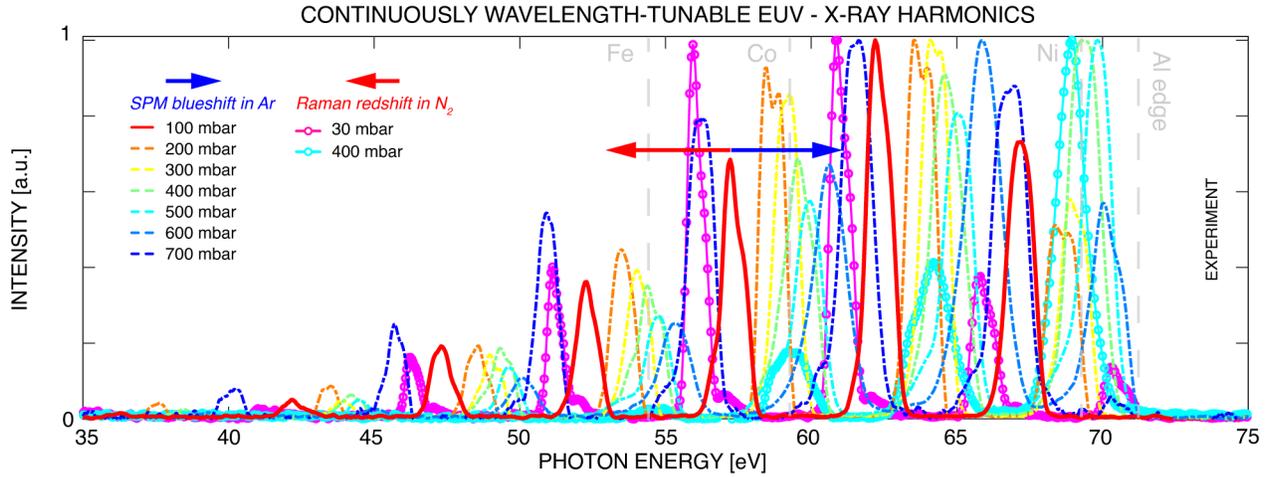

**Figure 3. Continuously red-shift and blue-shift wavelength-tunable EUV – X-ray high harmonics from He atoms, generated by a tunable VIS driving laser with central wavelengths spanning the broad range of $480 - 550\ nm$.** Blue-shift and red-shift wavelength-tunable EUV harmonics covering a complete spectral tunability of harmonic orders. The arrows indicate the shift direction when atomic $Ar$ (blue-shift) or molecular $N_2$ gas (red-shift) is used for spectral broadening (the color coding denotes the corresponding pressure).

In addition, recent theoretical studies uncovered that favorable attosecond-femtosecond Rabi oscillations might boost the conversion efficiency and prevent depletion of the ionized electronic state for even shorter EUV drivers [25]. In Fig. 3, we demonstrate the experimental prospect of generating high harmonics with maximum energy conversion and continuous tunability of the energy peaks with both blue and red shifts. This is enabled by the favorable interplay between the combination of optimized pressures in the waveguide and phase-matching angles of the single harmonic generating (SHG) crystal to tune the central laser wavelength. We note that, for an ultrabroad $1030\ nm$ spectrum, the HHG peaks have a very limited tunability by simply tuning the $BBO$ phase-matching parameters. In the process of tuning the gas pressure and, consequently, the carrier frequency of the SHG pulses, the pulses accumulate pressure-dependent varying phases. At first glance, the constant chromatic dispersion of the chirped mirrors appears to limit the pulse compression of the IR pulses. However, shorter visible pulses are beneficially obtained in the second harmonic generation process [13]. The temporal window for energy transfer between the fundamental and second harmonic pulses is limited by the group velocity walk-off and the BBO chromatic dispersion, which cause the pulses to reshape and separate. Nonetheless, the situation changes when the fundamental pulse tail contains substantial energy. Using chirped pulses, the rear of the fundamental pulse transfers power to the second harmonic peak at farther propagation distances, owing to the group-velocity mismatch [26-29]. This results in a sharp peak with shortened second harmonic pulse duration. Hence, the third ($TOD$) and higher dispersion orders act as a driving source for self-compression. In addition, the pulse chirping prevents further distortions of the pulse envelope during propagation in the dispersive media. Alternatively, an ultrashort short transform-limited fundamental pulse will require an extremely short crystal length to maintain a short pulse duration at the second harmonic, causing, in some cases, lower conversion efficiency.

In our experiments, the high-order harmonics are generated by tunable VIS pulses in the $480 - 550\ nm$ spectral range inside a gas cell filled with $He$ which is the unsurpassed medium for high cutoff and brightness upconversion due to its unique electronic structure providing high transparency and high ionization potentials [30]. The high harmonic beams are imaged by a gold-coated toroidal mirror and then spectrally separated by a blazed spherical grating with variable line spacing. The spectrally dispersed beams are detected by an X-ray CCD camera (Fig. 1). The blue and the red shifts of the VIS beam spectrum, realized inside of $Ar$ or $N_2$ − filled waveguide, allows for tuning of the high-order harmonic peaks by two photons or one odd harmonic in the EUV and soft-X-ray region. This exhibits a full tunability between two adjacent harmonics towards lower or higher energies while preserving the high harmonic brightness, which has not been demonstrated to date.

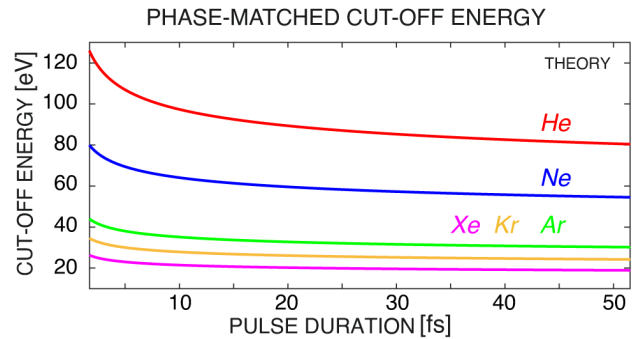

**Figure 4. Effective phase-matching limits for 1-30 laser cycles considering only the dispersion of atoms for a VIS driving laser at $515\ nm$.** Theoretical effective phase-matching limits for the cutoff high-order harmonic energy for visible driving laser with one to thirty cycles pulse duration are shown for several noble gases in matching colors. Only the dominant neutral atom dispersion is taken into account, while the contribution of the dispersion of ions which extends the cutoff of bright emission further, is excluded in this conservative estimate.

However, in experiments, full tunability is rarely needed. An energy scan or tuning of harmonic wavelength to a resonant absorption edge can be achieved by shifting adjacently positioned harmonic peaks either toward lower or higher energies, utilizing atomic or Raman-induced spectral shift in $N_2$ (Fig. 2) or other molecular gasses, i.e., air.

Furthermore, the harmonics are well separated by $4.8 - 5.2\ eV$ allowing for an easy selection of a resonant harmonic with X-ray mirrors, monochromators, or thin-film filters. In such a way, the source can emit light at a specific wavelength or continuously vary the wavelength in scanning experiments over a certain wavelength range. Additionally, the harmonics are generated with an excellent peak-to-valley contrast allowing direct experimental implementation in schemes where narrowband light is essential, e.g., in coherent diffractive magnetic imaging where off-resonance light can produce a large charge-scattering background.

Using visible driving wavelengths in the $480 - 550\ nm$ spectral range, the phase-matching conditions for bright coherent upconversion enable in our experiments, first, a wider temporal phase-matching window and, second, an effective phase-matched upconversion across the EUV spectral range and the soft X ray regime up to a phase-matching cutoff of $\sim 110\ eV$. The critical ionization level, which sets a limit to the maximum laser intensity, is favorably boosted due to the larger linear and nonlinear indices of refraction of noble gases for shorter-wavelength VIS-UV drivers as we approach the UV resonances of atoms. Furthermore, at higher intensities, the ions' refraction indices also become significant in the short-wavelength spectral regions. As a result, bright narrow linewidth harmonics can be generated due to the larger number of half cycles which contribute to the constructive addition of high harmonic fields. In addition, higher effective phase-matching cutoffs are allowed due to the increase in the limiting laser intensity. The broad bidirectional wavelength tunability of the driver leads to a smooth change of the high-order harmonic effective phase-matching conditions without substantial changes in phase-matching parameters, as shown in Figures 4 and 6.

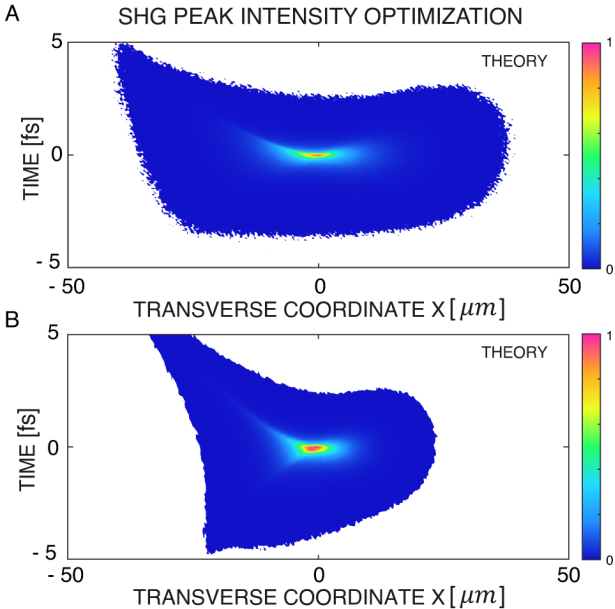

**Figure 5. Second harmonic peak intensity optimization of the VIS driving laser light.** Neural network optimization of flat wavefront second harmonic beam showing an optimal spatio-temporal intensity distribution at the focus for mirror radius of A) $R = 30\ cm$ and B) for the maximum intensity at $R \cong 18\ cm$.

In addition, to maximize the VIS laser intensity and, correspondingly, to extend the X-ray cutoff in this geometry, we use a very short $15\ cm$ focusing spherical mirror at a $3°$ incidence angle. These parameters are selected by a Neural-Network optimization using 3+1D Monte-Carlo ray tracing for the highest peak-intensity-product $1/\omega_x \omega_y \tau$ at the focus in restricted parameter space, where $\omega_x, \omega_y, \tau$ are the beam waists in $x$ and $y$ direction and the temporally broadened pulse duration due to space-time aberrations in the focusing geometry (Fig. 5), respectively [31, 32]. In addition, we vary the location where the peak-intensity product is estimated and effectively define the focus at the position of the circle of least confusion. The simulated SHG source spatial size has statistical Gaussian distribution with a standard deviation of $\sigma_{x,y} = 2.8\ mm$ and statistical divergence with a standard deviation of $\sigma_{\theta_x, \theta_y} = 100\ \mu rad$, corresponding to an M-squared value of $M^2 \sim 3$. The statistical approach provides an efficient way to evaluate the beam parameters, both spatial and temporal, at the desired position with a relatively small number of rays. Our calculations indicate that for the selected beam waist and divergence, spherical mirrors down to nearly $10\ cm$ of focal length can be used to achieve tight focusing before the spatio-temporal aberrations cease the intensity growth. In this configuration, we experimentally observe harmonics up to $>110\ eV$ emitted from neutral He atoms (Fig. 6). No emission from He ions is observed, as indicated by the strong-field approximation (SFA) calculations in Fig. 6 (C). In the future, increasing the peak intensity of short-wavelength UV-VIS driving laser beams can effectively phase-match harmonics in the "water window" at $285 - 420\ eV$, where water is transparent, hence allowing for in vivo dynamic studies of artificial or living specimens, and beyond.

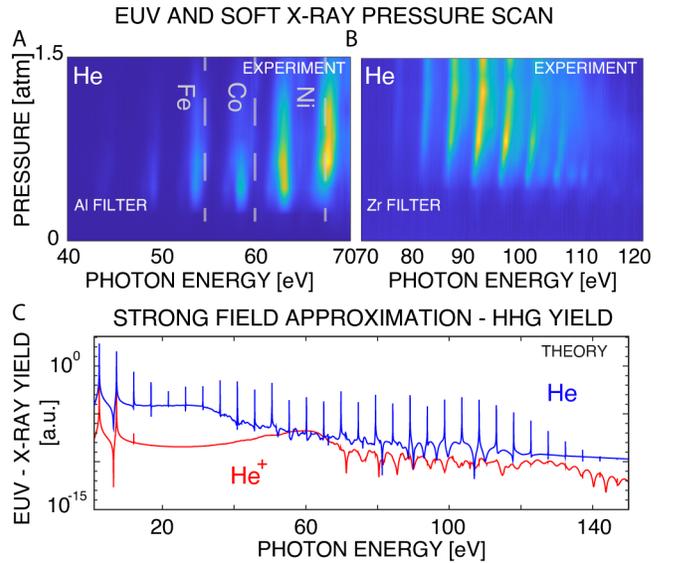

**Figure 6. Continuously wavelength-tunable EUV – X-ray high harmonics from He atoms, generated by a tunable VIS driving laser with wavelengths spanning the broad range of $480 - 550\ nm$.** A) and B) Experimental backing-pressure dependence of the high harmonics from $He$, extending into the soft-X-ray region for $Ar$-filled waveguide at $400\ mbar$. C) Single-atom estimate of the harmonic yield from neutral He atoms and He$^+$ ions at an intensity of $10^{15}\ W/cm^2$.

**Conclusion.**

In summary, our results demonstrate a highly versatile source of continuous wavelength-tunable bright harmonics, resonantly covering the magnetic *N* and *M* absorption edges of ferromagnetic elements: $Fe \sim 54\ eV$, $Co \sim 60\ eV$, $Ni \sim 68\ eV$, $Gd \sim 20\ eV$, and $145\ eV$. As indicated by calculations, the harmonics emerge as a train of sub-300 attosecond pulses in the time domain, allowing for ultrafast pump-probe multidimensional (4+1D) imaging and XAFS spectroscopies in the near future, where the additional effective dimension is the material identity. Moreover, we generate continuously wavelength-tunable, visible beams with ultrashort pulse duration down to $8.4\ fs$, owing to direct nonlinear self-compression, further enhancing the potential applications of this source for optical spectroscopies.

**Funding.** H2020 European Research Council (100010663), XSTREAM-716950, Alfred P. Sloan Foundation (100000879) FG-2018-10892

**Acknowledgments.** TP acknowledges funding from the European Research Council (ERC) under the European Union's Horizon 2020 research and innovation program (grant agreement XSTREAM-716950) and from the Alfred P. Sloan Foundation (FG-2018-10892).

**Disclosures.** The authors declare no conflict of interest.

**Data availability.** Data underlying the results are presented in the plotted graphs.